\documentclass[twocolumn,pre,showpacs,superscriptaddress]{revtex4}  

\usepackage{graphicx}
\usepackage{amssymb}

\begin{document}

\title{Characterization of the stretched exponential trap-time
distributions in one-dimensional coupled map lattices }

\author{S. I. Simdyankin} 

\author{Normand Mousseau} 

\affiliation{D\'epartement de physique and Centre de recherche en physique
et technologie des couches minces, Universit\'e de Montr\'eal,
C.P. 6128, succ. Centre-ville, Montr\'eal (Qu\'ebec) H3C 3J7, Canada}

\author{E. R. Hunt} 

\affiliation{Department of Physics and Astronomy and CMSS, Ohio
University, Athens, OH 45701, USA}

\date{\today}

\begin{abstract}
Stretched exponential distributions and relaxation responses are
encountered in a wide range of physical systems such as glasses,
polymers and spin glasses.
As found recently, this type of behavior occurs also for the
distribution function of certain trap time in a number of coupled
dynamical systems.
We analyze a one-dimensional mathematical model of coupled chaotic
oscillators which reproduces an experimental set-up of coupled
diode-resonators and identify the necessary ingredients for stretched
exponential distributions.
\end{abstract}


\maketitle

%-------------------------------------------------------------------------
%			Main text
%-------------------------------------------------------------------------
%
%
\section{Introduction}
\label{sec:intro}

The decay of certain quantities characteristic of many complex systems
such as glasses~\cite{Phillips_96}, spin glasses~\cite{Campbell_85},
quasi-crystals~\cite{Dzugutov_95}, trapping
models~\cite{Donsker_75,Grassberger_82,Barkema_2001}, coupled
non-linear systems~\cite{Hunt_2002,Roy_97,Pruthi_95},
turbulence~\cite{Frisch_97,Gamba_99} and
others~\cite{Weibull_51,Laherrere_98,Jonscher_99} is often described 
by a stretched-exponential -- or Kohlrausch -- functional form:
\begin{equation}
\phi(t) = \exp \! \left[ - (t/\tau)^{\beta} \right], \; t \ge 0
\label{strexp}
\end{equation}
with $0 < \beta < 1 $~\footnote{By analogy, the case $1< \beta < 2$ is
often called ``stretched Gaussian'' (see e.g.
Refs.~\onlinecite{Kaplan_96,vanZon_02}).}.
Although Eq.~(\ref{strexp}) provides a good fit to a wide range of
experimental and numerical results, in many cases these can also be fitted by
power laws with comparable accuracy;
while most experimental set-ups can span many decades in
time, few have achieved more than 2 or 3 decades in $\phi(t)$.
The Kohlrausch form can also be reproduced explicitely by a few
theoretical models (see, e.g.,
Refs.~\onlinecite{Donsker_75,Grassberger_82,Campbell_85,Gielis_95,Garrahan_2002})
on the basis of various assumptions, but it is still unclear whether
there can be a unifying solid theoretical justification for it.

Recently, Hunt, Gade and Mousseau~\cite{Hunt_2002} found that
stretched exponentials could fit experimental distributions of trap
time~--- the time a system spends in an uninterrupted state with
temporal period two, in a one-dimensional network of coupled diode
resonators~--- over more than 6 decades in the distribution.
Numerical models, based on the set-up, could expand this fitting over
about 10 orders of magnitude, strongly suggesting that dynamics with
underlying stretched exponential distributions could be universal in
coupled chaotic systems and ruling out a power-law or any other
standard fit.

In this paper, we revisit one such numerical model and provide a
detailed characterization of the dynamics of this network as a
function of system size and parameters, providing elements of
explanation regarding the origin of stretched exponential
distributions in this system.
In particular, we show that 
(1) size effects are important only for relatively small systems;
(2) the natural invariant density, $\rho(x)$, generated by typical
orbits has well-defined structure that is self-organized;
(3) the structure of $\rho(x)$ alone cannot lead to
stretched-exponential distributions~--- the dynamical spatial
organization is essential for stabilizing the periodic orbits.

This paper is organized as follows: In the next section, we introduce
the mathematical model: a one-dimensional coupled map lattice.
Section~\ref{sec:background} discusses the background for the
trap-time distributions.  In section~\ref{sec:results}, the results
from the simulation are presented, to be discussed in section
\ref{sec:discussion}. We summarize our conclusions in
Sect.~\ref{sec:conclusion}.

\section{Model} 
\label{sec:model}

Our model is a one-dimensional chain of $N$ diffusively coupled nonlinear
deterministic maps, $f(x)$, with a coupling constant $\alpha$ and periodic
boundary conditions. The interaction is totalistic and involves only the
nearest neighbors.  The time evolution of this system is discrete and
is described by an iterative equation: 
\begin{equation}
x_n(t+1) = (1-\alpha) f[x_n(t)] + 
                  \frac{\alpha}{2} \left\{ f[x_{n-1}(t)] +
                                           f[x_{n+1}(t)] \right\}.
\label{cml} 
\end{equation} 
which, given initial conditions $x_n(0)$ for each site
$n=1,2,\dots,N$, generates a time series 
$\{x_n(t)\}$, for integer values of the time index
$t=0,1,\dots$.
This model offers a much simplified version of the 1D array of diode
resonators studied in
Refs.~\onlinecite{Johnson_96,Locher_98,Locher_98b,Locher_00}. Although
the individual diode resonators are best described by an inertial
equation~\cite{Rollins_84}, it was shown recently that Eq.~\ref{cml}
captures the dynamics in the regime of interest here~\cite{Hunt_2002}.

The diffusive nature of the coupling in Eq.~(\ref{cml}) becomes more
apparent if this equation is rewritten in the following form
\begin{eqnarray}
x_n(t+1) & = & f[x_n(t)] + \nonumber \\
         &   & \frac{\alpha}{2} \left\{ f[x_{n-1}(t)] - 
               2f[x_n(t)]+ f[x_{n+1}(t)] \right\} \nonumber \\
         & = & f[x_n(t)] + \frac{\alpha}{2} D_+D_-f[x_n(t)] 
\label{cmlD} 
\end{eqnarray} 
where the central difference operator $D_+D_-f(x_n)$ is the discrete
Laplacian of $f(x_n)$ on a one-dimensional grid with unit spacing. In
the limit $N\to\infty$ the discrete Laplacian can be substituted by
its continuous counterpart and Eq.~(\ref{cmlD}) becomes a nonlinear
analogue of a difference-differential diffusion equation, see
e.g., Ref.~\onlinecite{Elmer_2001}.
For a large $N$, $1/r$ can be thought of as a ``viscosity'' and
$\alpha$ as a ``diffusivity'' with their usual relation to the
``temperature.''

We use the logistic map $f(x) = rx(1-x)$,
Ref.~\onlinecite{Ott_CDS}, in order to describe the dynamics of the
basic chaotic elements.
Although this differs from the form studied in
Ref.~\onlinecite{Hunt_2002}, $g(y)=1-a y^2$, the results are
unaffected; the two maps are conjugate, related by a simple algebraic 
transformation: $x = \sqrt{a}y/\sqrt{r}+1/2$, $g(y) = f[x(y)]-r/4+1$. 
We choose the logistic map because it is more studied in the literature.
In the simulations presented here, we fix the value of the coupling
constant to $\alpha = 0.25$, and vary $N$ and $r$, which we refer to
as the nonlinearity parameter in the following. 

Initial values of $x_n(0)$ are taken from a random distribution
in the $[0,1[$ interval. Runs are then iterated for a few hundred
thousands steps in order to avoid any transient effect before
starting the accumulation of data.
All simulations presented here are done on one-dimensional arrays with
variable length and periodic-boundary conditions. 
Statistics are generally accumulated over 10 million to 10 billion
time steps.

Following the experiment \cite{Hunt_2002}, the analysis of the
dynamics is done using coarse-grained variables defined by
\begin{equation}
\sigma_n(t) = \mathrm{sign}[x_n(t)-x_\mathrm{thr}], \; t=0,2,4,\dots
\label{renew}
\end{equation}
where the quantity $x_n(t)$ is defined in Eq.~(\ref{cml}) and
$x_\mathrm{thr}$ is a certain threshold value. The results presented
here are not very sensitive to the value of the threshold. For
simplicity, we select the value of the unstable fixed point, $x_* = 1
- 1/r$, for the single map.  This coarse-graining reduces the problem
from continuous to two-state.

The basic dynamics of the coupled oscillator being period two, the analysis
is also done only over even (or odd) time steps. 

\section{Background: the distribution of trap time}
\label{sec:background}

\begin{figure} % 
\centerline{\includegraphics[width=8.5cm]{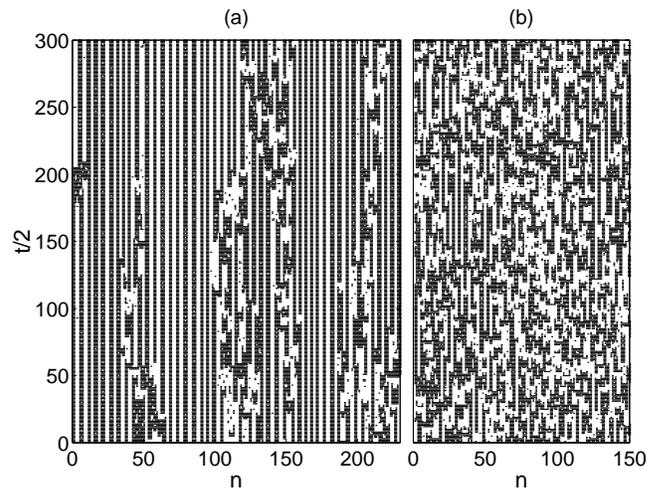}}
\caption{Coarse-grained time evolution of the coupled logistic map.
The coarse-grained variable $\sigma_n(t)$ in a stroboscopic
representation : for every other iteration $t$ a dot at $(n,t)$
corresponds to $\sigma_n(t)=+1$ and the blank space signifies
$\sigma_n(t)=-1$. $N=1000$ and (a) $r=3.83$, (b) $r=3.8888$.  }
\label{fig:bin_map}
\end{figure}

We are interested in the statistical distribution of trap time in a
coarse-grained state space of a one-dimensional chain of coupled
nonlinear maps, as shown in Fig.~\ref{fig:bin_map}.
This quantity is formally equivalent to the distribution of time
intervals between zero crossings of renewal processes such as random
walks~\cite{Godreche_01} and has the advantage that it can be measured
experimentally and numerically to a high degree of
accuracy for this system~\cite{Hunt_2002}.

Generally, however, experimentally measurable relaxation responses are
mathematically described by auto-correlation functions of certain
dynamical variables.
For example, the inverse Fourier transform of the dynamical structure
factor, the intermediate scattering function, can be calculated as time
auto-correlation function of the microscopic density distribution of
particles in a material~\cite{Hansen_TSL}.

The relation between the auto-correlation function $C(t) = \langle
\sigma(t')\sigma(t'+t) \rangle_{t'}$ of a renewal process $\sigma(t)$
---defined analogously to Eq.~(\ref{renew})~--- and the distribution
of trap time of this process was recently studied for a range of
distributions~\cite{Godreche_01}.
Using this framework, it is possible to show that $C(t)$
corresponding to the stretched exponential distribution of occupation
time is also stretched exponential at long time albeit with a
different stretching exponent $\beta$.  

These results are only valid when the traps are
uncorrelated in time which is the case for the systems studied here.
The details of this work will be presented
elsewhere~\cite{Simdyankin_2002_PRL}.

This establishes a direct relation between the distribution of trap
time, discussed in this paper, and the more standard auto-correlation
function, measured in a number of experiments on glasses and other
complex systems.

\section{Results}
\label{sec:results}

Before discussing the possible origins for 
the stretched exponential distributions in our model, it is necessary
to offer a characterization of its dynamics as a function of the
parameters of the model.

First, we consider the size effects on the dynamics of the
network, studying experimental set-up varying in length between 15 and
256 oscillators, and simulated arrays of up to tens of thousands of
sites. 

Then we investigate the dynamics of the system as a function of the
nonlinearity parameter, $r$, which brings the system through a
series of dynamical changes from stable periodic orbits to full chaos.
In particular, it is important to assess the parts of the
parameter space where the stretched exponential distributions can be
observed.

Once the basic phase diagram is established, we discuss the building up of the
dynamics on a single site embedded in the network as well as the
spatial structure associated with the stretched exponential
distribution.

\subsection{Dependence on $N$}
\label{subsec:N}

\begin{figure} 
\centerline{\includegraphics[width=8.5cm]{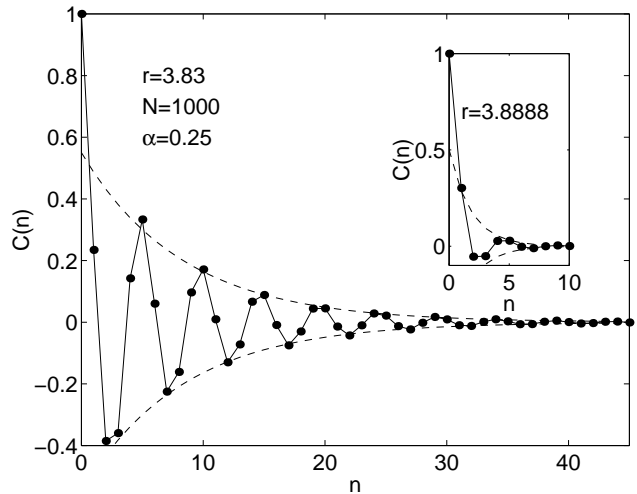}}
\caption{Spatial correlation function $C(n)=\langle \sigma_{n'}(t)
\sigma_{n+n'}(t) \rangle_{n',t}$.  Dots correspond to the numerical
data. Solid line interpolates numerical points and is shown as a guide
to the eye. Dashed lines show the exponential decay of oscillations:
$\pm 0.55 \exp(-n/8.25)$. Inset also shows $C(n)$, but calculated for
a different value of $r$. The amplitude of oscillations (dashed lines)
in this case decays as $\pm 0.5 \exp(-n/1.8)$.  }
\label{fig:spatial}
\end{figure}

As is seen in Fig.~\ref{fig:bin_map}, the traps are directly
associated with a periodic spatial organization, which is controlled
by the coupling $\alpha$. 

As $\alpha$ is increased, the spatial organization goes
through two rapid transitions, showing qualitatively different
features:
For $\alpha < 0.1$, the lattice tends to follow a period-two spatial
organization. 
For intermediate values of $0.1 < \alpha < 0.19$, the lattice
immediately freezes into a period-two state in both space and time,
for all values of the driving parameter, $r$. 
Above this threshold, $\alpha>0.19$, the dynamics becomes stochastic
again while dominant spatial period goes to 4 and even longer for
large $\alpha$.
In each of these phases, the variation of $\alpha$ affects only
minimally the spatial structure of the phase.

In spite of the evident organization seen in
Fig.~\ref{fig:bin_map}, the spatial correlation is short-range. 
Figure ~\ref{fig:spatial} shows that spatial correlations vanish
exponentially fast with a typical length scale between about 2 and 8,
i.e., the only static spatial correlation appearing in the system is
directly associated with the short-range organization.

\begin{figure} 
\centerline{\includegraphics[width=8.5cm]{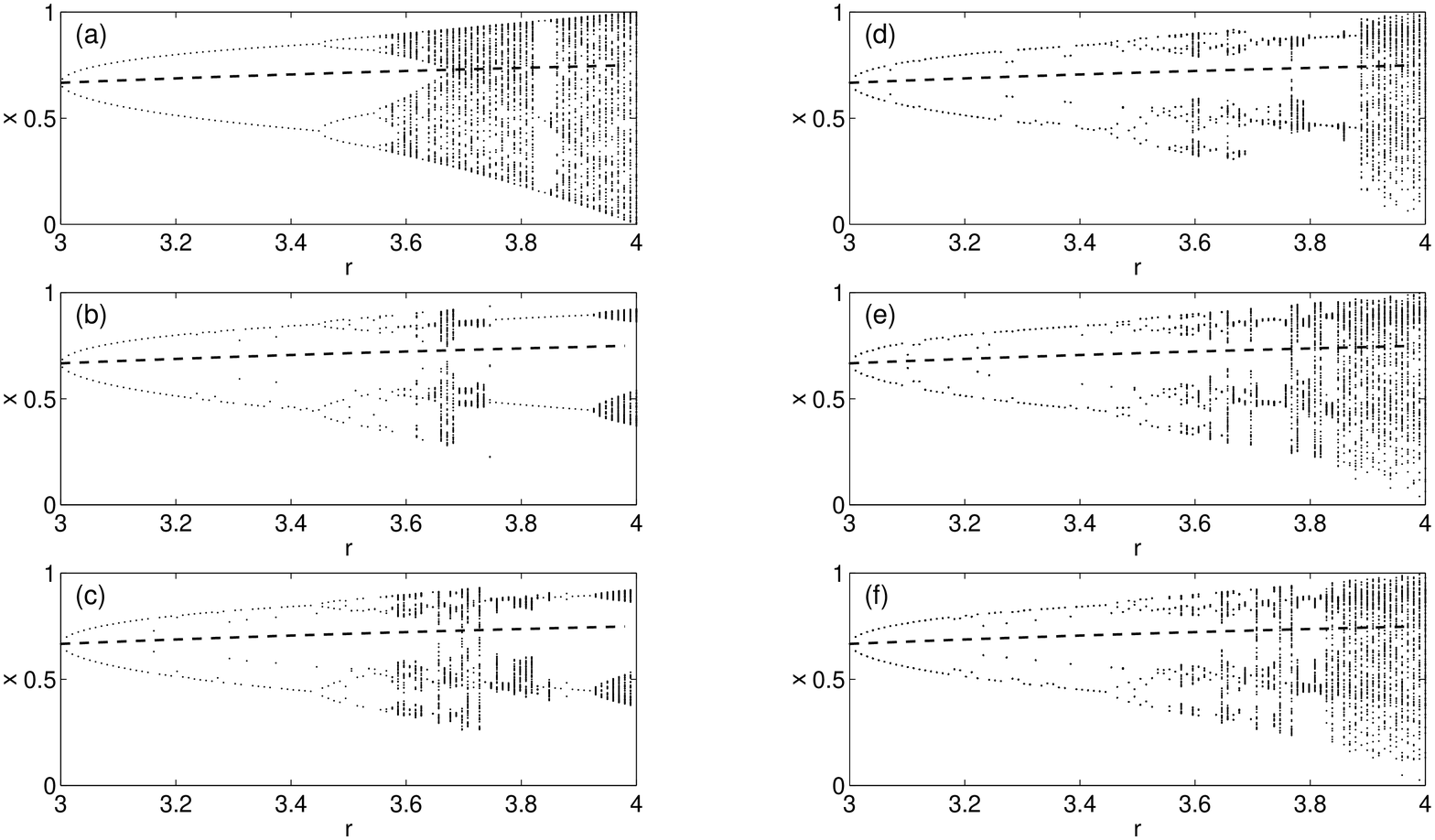}}
\caption{Bifurcation diagrams for (a) an isolated map, (b) $N=8$, (c)
$N=16$, (d) $N=32$, (e) $N=1024$, and (f) $N=10000$. Dashed line in
all panels shows the unstable orbit $x_* = 1 - 1/r$. }
\label{bifdiag}
\end{figure}

This sort-range correlation implies that size effects should be very
limited.
Fig.~\ref{bifdiag} shows the lattice-size dependence on the
bifurcation diagram, for a single site on the lattice.

\begin{figure} 
\centerline{\includegraphics[width=8.5cm]{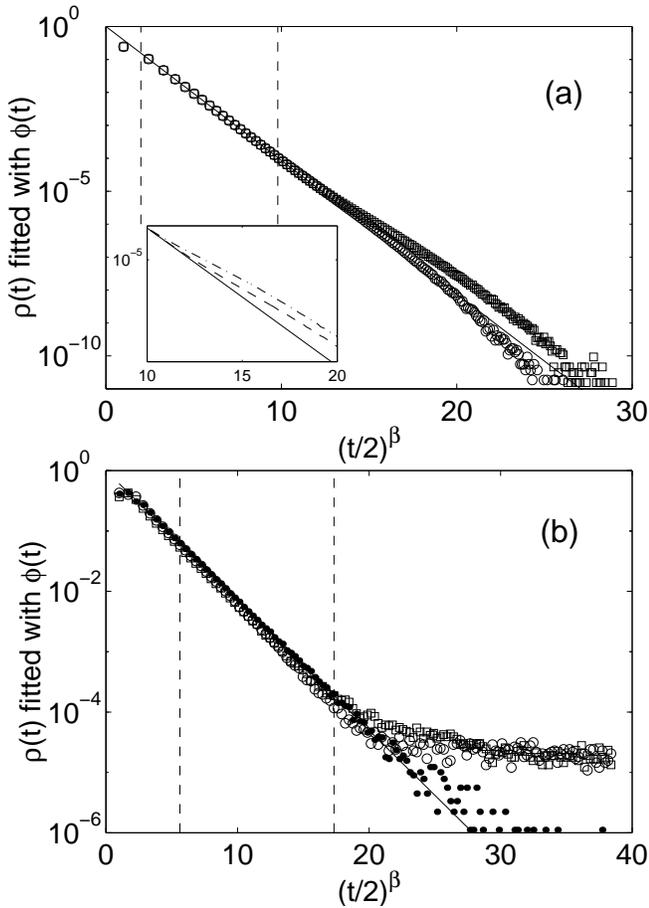}}
\caption{(a) Distribution of trap time for $r=3.8888$. $N=21$
(squares: set of data points lying above the solid line) and $N=19$
(open circles: the set of data points lying below the solid line). The
solid line corresponds to the stretched exponential fit
Eq.~(\ref{strexp}) with $\beta=0.50\pm0.05$ and $\tau=3.0\pm0.5$.  In
inset, we show the data for $N=17$ (dash-dotted line), $N=21$ (dashed
line) and the fit (solid line). (b) Distribution of trap time for the
experimental set-up (squares: $N=21$, dots: $N=20$, open circles:
$N=19$). The stretched exponential fit is with $\beta=0.75\pm0.05$ and
$\tau=5.0\pm0.5$.  Vertical lines in both panels demarcate the
interval over which the fitting procedure was performed.}
\label{fig:Ns}
\end{figure}

For small lattices, $1 < N < 32$, the dynamics depends on whether $N$
is even or odd.
For even $N$, arrays synchronize rapidly while the odd sizes
continue to display chaotic trajectories.
For example, the bifurcation diagram changes qualitatively as one goes
from a single isolated site to a chain of 8 or 16 oscillators: the
chaotic phase disappears totally and the system remains periodic in
the coarse-grained state space until $r=4.0$ [see
Fig.~\ref{bifdiag}(b) and (c)].
While isolated oscillators produce an exponential trap-time distribution,
this distribution tends to the stretched exponential form at short
times for lattices with odd $N$ as small as 15.
Size effects are still present for these lattices, however, and the
long-time distribution diverges from the stretched exponential [see
Fig.~\ref{fig:Ns} (a)].
Interestingly, the sign of the deviation for the stretched exponential
oscillates as the array is increased in size two by two, indicating a
certain spatial frustration in these small systems.

For even $N$ that is not a multiple of four, the mismatch with the
periodicity results in the presence of inclusions of stable defects,
with four sites at a row in the same band.
These defects can also be present and stable in lattices where $N$ is
a multiple of four.

As the number of elements reaches 32, orbits corresponding to larger
values of $r_{\mathrm{min}} \le r \le 4$ become chaotic and the
bifurcation diagram approaches that of an
infinite lattice.  $r_{\mathrm{min}}$ continues to decrease with
growing number of sites, and is well converged for a lattice of a few
hundred oscillators.
As shown in Fig. ~\ref{fig:Ns} (b) the same trend is seen
experimentally, although the inherent disorders helps to decrease the
finite size effects.
%
%MORE ON EXPERIMENTS...

Even though the bifurcation diagrams for an isolated map and the
infinite lattice appear similar, their respective orbits are
qualitatively different for most values of $r$. In particular, most of
the chaotic region of the latter shows a stretched exponential
trap-time distribution.

\subsection{Dependence on $r$}
\label{subsec:r}

\begin{figure} % 
\centerline{\includegraphics[width=8cm]{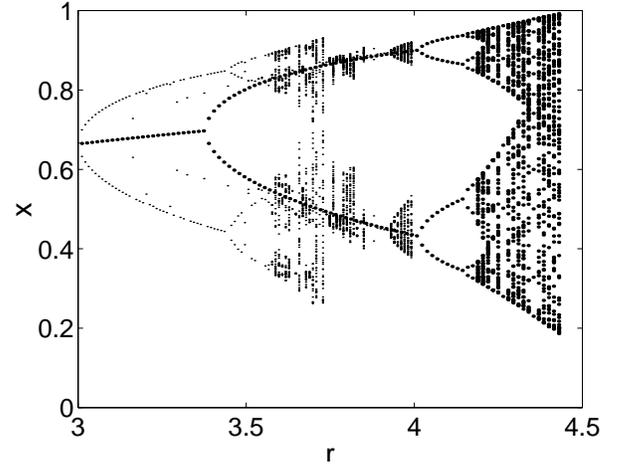}}
\caption{Larger dots: bifurcation diagram of the map $x(t+1) = 0.75 \,
f[x(t)]+0.1625$.  Smaller dots: the same as in Fig.~\ref{bifdiag}
(c). Although $\rho(a)$ [dashed-dotted line in
Fig.~\ref{fig:pert_distr} (c)] corresponding to the bifurcation
diagram depicted by the smaller dots is not a $\delta$-function, it is
narrow enough so that it creates a period-two attractor in the coarse
grained phase space of the couple map chain.}
\label{fig:bdonepconst}
\end{figure}

Fig.~\ref{bifdiag} also indicates the effect of $r$ on the dynamics of
a single site in a chain.
The coupling stabilizes the orbits, reducing significantly the size of
the chaotic region.
Its effect is to shift the bifurcation diagram to the right, moving
the dynamics towards periodic orbits as is shown in
Fig.~\ref{fig:bdonepconst} (Fig.~\ref{fig:bdonepconst} will be discussed 
in more detail in Sect.~\ref{subsec:single}).

\begin{figure} % 
\centerline{\includegraphics[width=8.5cm]{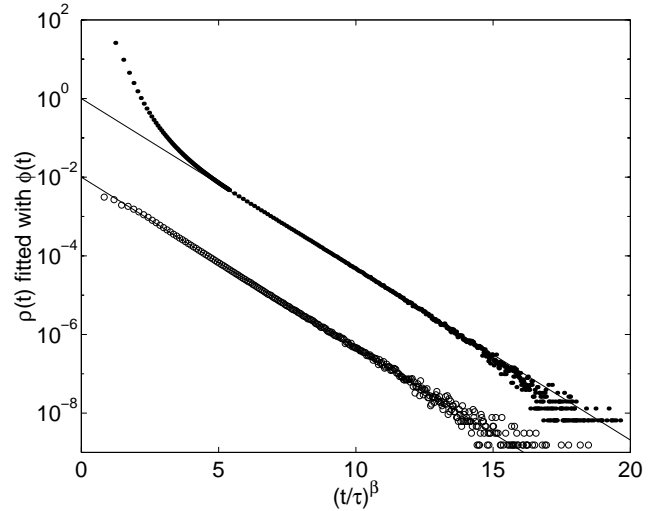}}
\caption{Trap-time distributions fitted with stretched exponentials.
Dots correspond to $N=1000$, $r=3.83$ and $\beta=0.33\pm0.05$,
$\tau=11.5\pm0.5$. Open circles correspond to $N=64$, $r=3.8888$ and
$\beta=0.50\pm0.05$, $\tau=2.9\pm0.5$.  The solid lines show the best
fit to the stretched exponential function $\phi(t)$.}
\label{fig:rnewhist}
\end{figure}

The stretched exponential behavior appears, for a large enough
lattice, around $r_0=3.83$. At this threshold value, the stretching
exponent $\beta \approx 0.33$, obtained from fitting the long-time
part of the trap distribution (Fig.~\ref{fig:rnewhist}) according to
Eq.~(\ref{strexp}). $0.33$ is the lowest value of $\beta$ we could
observe for the explored regions of the parameter space of the model.
Below this threshold, chaotic windows are interspersed with
periodic ones.  Trajectories in these narrow windows tend to fall
erratically onto neighboring periodic orbits, generating multiple-step
trap distributions with no clear overall time behavior.

As seen in Fig.~\ref{fig:rnewhist}, the value of $\beta$ increases
with $r$. For $r=3.8888$, the trap-time distribution is well fitted by
a stretched exponential with $\beta \approx 0.5$, while at $r=4$, the
full chaos limit for an isolated logistic map, a fit to $\rho(t)$
gives $\beta=0.70\pm0.05$ and $\tau = 3.0\pm0.5$.

Because of the short spatial correlation and the excellent quality of
the simulation data, the exact value of $\beta$ is well defined:
increasing the lattice size from 64 to 1000 sites leaves the
trap-distribution essentially unchanged. Moreover, the same values of
$\beta$ are obtained, within the error bars, by changing the length of
the interval over which the fitting procedure is performed.

These numbers are also in reasonable agreement with experimental results
which show $\beta$ varying from $0.10\pm 0.05$ to $0.95\pm 0.05$. This
wider range for the experiment is probably caused by the
presence of site disorder.

This confirms that the best functional form for the simulation data is
given by the Kohlrausch function.

\subsection{Single site dynamics.}
\label{subsec:single}

The iterative rule Eq.~(\ref{cml}) can be viewed as an equation
describing the dynamics of a single element in the presence of an
external additive perturbation $a(t)$:
\begin{equation}
x(t+1) = (1-\alpha) f[x(t)] + \alpha \, a(t) 
\label{eq:pert_map} 
\end{equation} 
where $a(t) = \{f[x_{n-1}(t)] + f[x_{n+1}(t)]\}/2$.
For simplicity, since all sites are statistically identical, we drop
the subscript $n$ in Eq.~(\ref{eq:pert_map}).

This form allows us to concentrate on the impact, at the single-site level, of
the rest of the network and to try to identify the essential elements
for a stretched exponential dynamics. 
For this purpose, it is convenient to use the natural invariant
density $\rho(x)$ generated by typical orbits $\{x(t)\}$,
$t=1,2,\dots$ of the map in Eq.~(\ref{eq:pert_map}).
The notion of the natural invariant density is widely used in the
studies of chaos \cite{Ott_CDS}.
The function $\rho(x)$ is defined so that for any interval
$[x,x+dx]\in[0,1]$ the fraction of the time typical orbits spend in
this interval is $\rho(x)dx$.
In the same way one can define the density $\rho(a)$ for the
perturbation $a(t)$ in Eq.~(\ref{eq:pert_map}).

\begin{figure*} % 
\centerline{\includegraphics[width=16cm]{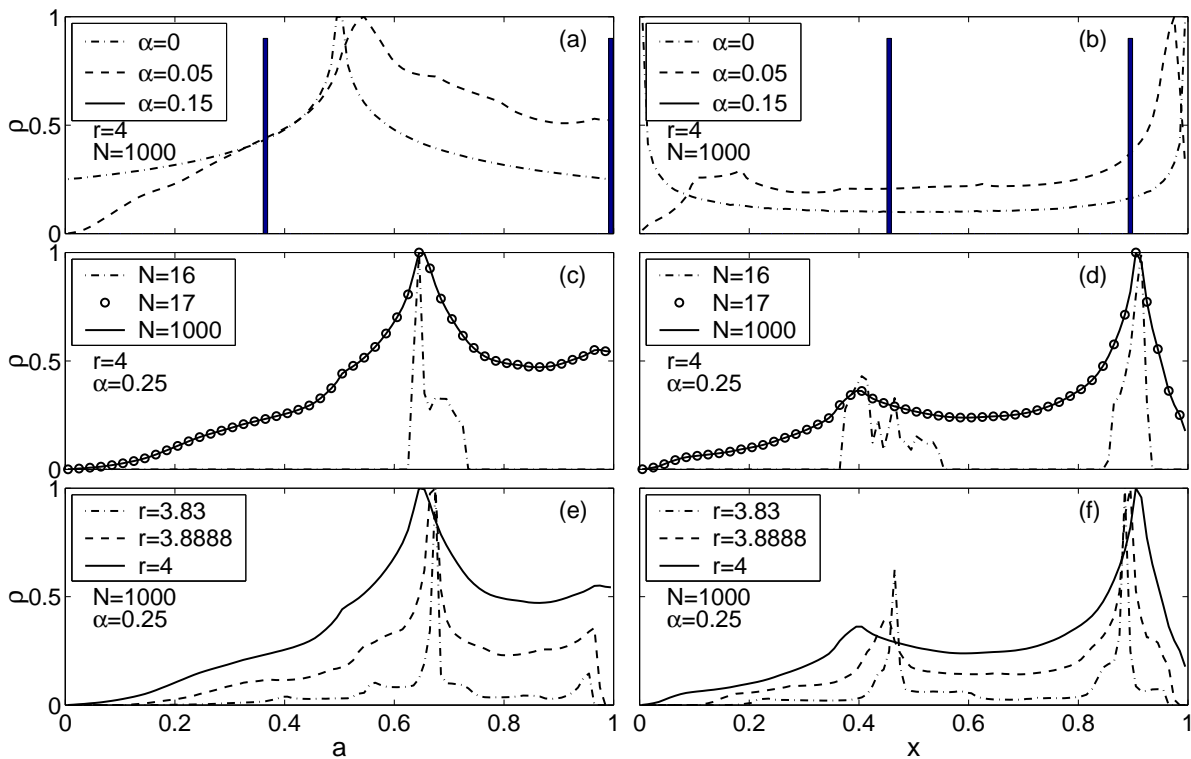}}
\caption{Densities $\rho(a)$ and $\rho(x)$ of $a(t) =
\{f[x_{n-1}(t)] + f[x_{n+1}(t)]\}/2$ [panels (a),(c),(e)] and x(t)
[panels (b), (d), (f)] from Eq.~(\ref{eq:pert_map}) normalized by the
maximal value. Note that each row of panels corresponds to the same
sets of parameters.}
\label{fig:pert_distr}
\end{figure*}

Fig.~\ref{fig:pert_distr} shows the behavior of $\rho(a)$ and
$\rho(x)$ for a few sets of the parameters $\alpha$, $N$, and $r$.
In the case of an isolated map ($\alpha=0$) and in the chaotic regime,
$a(t)$ is the sum of two identically distributed independent random
variables~--- $\rho(a)$ is shown by the dash-dotted line in
Fig.~\ref{fig:pert_distr} (b).
According to the central limit theorem, if this sum contained much
more than two term, $\rho(a)$ would very well agree with the normal
(Gaussian) distribution.
However, even with only two terms, $\rho(a)$ exhibits a maximum at
$a=1/2$ [the dash-dotted line in Fig.~\ref{fig:pert_distr} (a)].

The presence of one or few peaks in $\rho(a)$ makes the orbits
$\{x(t)\}$ intermittent for some values of the nonlinearity parameter
$r$.
This intermittency means that the (coarse-grained) orbit
$\{\sigma(t)\}$, see Eq.~(\ref{renew}), stays periodic during some
time, becomes chaotic for a certain while, and goes periodic again.
In order to demonstrate how a periodic attractor develops in a coupled
system at a value of $r$ for which the isolated map produces a chaotic
orbit, we can imagine a density $\rho(a)=\delta(a-a_0)$,
where $\delta(a)$ is the delta function and $a_0\in[0,1]$ is a
constant.
In this case, Eq.~(\ref{eq:pert_map}) becomes
$x(t+1)=(1-\alpha)f[x(t)]+\alpha \, a_0=c_1 f[x(t)]+c_2$, which is the
iterative equation for the isolated map scaled by a constant $c_1$ and
shifted by another constant $c_2$.
This manipulation with the isolated map results in a bifurcation
diagram translated with respect to the original one
[Fig.~\ref{bifdiag} (a)] while preserving all its qualitative
features, see Fig.~\ref{fig:bdonepconst}.
In particular, the period-two attractors of the modified map overlap
with chaotic orbits of the original map.

%The dashed curve in Fig.~\ref{fig:pert_distr} (a) shows how the
%structure of $\rho(a)$ develops for a small coupling $\alpha$ and the
%corresponding curve in Fig.~\ref{fig:pert_distr} (b) exhibits two
%side maxima indicating the presence of intermittent period-two
%dynamics.

As discussed above, in the interval $0.1 < \alpha < 0.19$, the lattice
falls into period-two spatial and temporal patterns, i.e., each two
neighboring sites oscillate out of phase and each site alternates
between two values $x^{(1)} < x^{(2)}$, as indicated in
Fig.~\ref{fig:pert_distr} (a) and (b).

For larger values of $\alpha$, the natural invariant density becomes
more complex. 
When the lattice size is a small even number, the spatial pattern is
periodic with the period which is equal to four (two sites ``up''
followed by two sites ``down'') and each site again undergoes a
period-two dynamics [Fig.~\ref{fig:pert_distr} (c) and (d)] which 
is manifested by the peaks in $\rho(x)$. 
For $N=16$, there are narrow peaks in $\rho(a)$ and $\rho(x)$.
These peak broaden and merge rapidly, however, for large $N$ and the
system displays strong sign of chaoticity.

%The densities $\rho(a)$ and $\rho(x)$ are practically
%indistinguishable if $N$ is not a multiple of four [dashed lines in
%Fig.~\ref{fig:pert_distr}, (c) and (d)] or it is sufficiently large so
%that the system is close to its continuum limit, see
%Sec.~\ref{sec:model}, [solid lines in Fig.~\ref{fig:pert_distr}, (c)
%and (d) which cover the dashed ones].
%
%The intermittent dynamics in this regime is manifested by the
%backgrounds underlying the maxima in $\rho(a)$ and $\rho(x)$.

This chaoticity can be controlled by $r$ and we see that the orbits
spend more and more time in period-two attractors relative to the
chaotic ones as the nonlinearity parameter is decreased, relating
directly to a change in the value of the stretching exponent $\beta$
[Fig.~\ref{fig:pert_distr} (e) and (f)].

Although providing a clearer picture of the dynamics, knowledge of the
invariant density is not sufficient to reproduce the stretched
exponential dynamics. 
For example, choosing randomly the values of $a(t)$ in
Eq.~(\ref{eq:pert_map}) according to a prescribed $\rho(a)$ generates
only an exponential trap-time distribution.
Variations on this theme, including the introduction of a two-step
distribution, which favors staying in the period-two phase once in it,
also fail to give a stretched exponential distribution of trap times.
Spatial correlations, even though short range, are essential to induce
a stable and self-organized stretched exponential distribution.

\subsection{Stability of spatial period-4 structure.}
\label{sec:stability}

Fig.~\ref{fig:bin_map} plots the time sequence for $\sigma_n(t)$ on a
small region of a 1000-site chain for $r=3.83$ ($\beta \approx 0.33$) and
$r=3.8888$ ($\beta \approx 0.5$). These show a remarkable range in the
trap size: the larger traps of the first one can be up to 250 times
longer than those at $r=3.8888$, and corresponding to a time-scale
30,000 times larger than the basic time-step. This time scale is
similar to that observed experimentally~\cite{Hunt_2002}. In spite of
these very stable traps, for long enough time, all sites display an
identical dynamics and no region of the lattice remains frozen.

Looking at the same figure, we also see that, {\it for a given $r$},
the trap width appears to be uncorrelated with its length: long traps
remain narrow and are composed of ordered domains with several
occurrences of period 4, 5 or 6.
While the period-4 domains can remain totally stable for a long time,
the middle site of the 3-up or 3-down segment  in period 5 and period
6 domains show some instability. They must therefore be considered
more as a defect in the stable phase than as an additional spatial
structure.  
Even so, these larger basic domains, that occur mostly for the lower
values of $r$, are responsible for the apparent spatial periodicity of
5 in the spatial correlation function at $r=3.83$ (see
Fig.~\ref{fig:spatial}).

\begin{figure} % 
\centerline{\includegraphics[width=8.5cm]{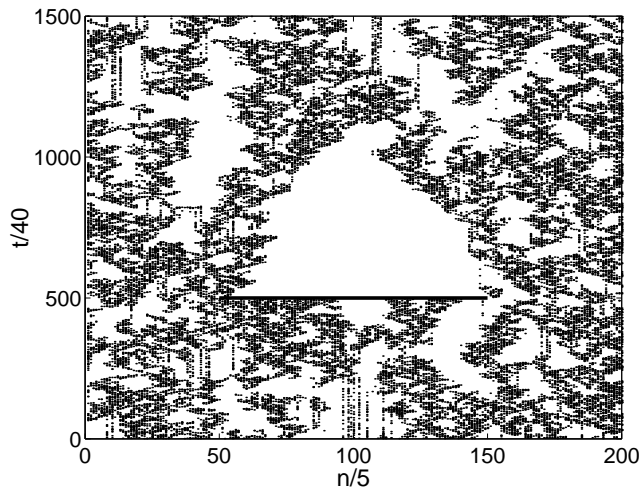}}
\caption{Space-time diagram for $N=1000$ and $r=3.83$.  Dots
correspond to trap boundaries~--- points in time when a site
interrupts its temporal period-2 dynamics. Every fifth site and every
40th iteration are shown. The heavy horizontal line shows the region
which was set into spatial period-4 state at a certain moment in
time. Note that thus induced spatial period-4 domain does not have any
bursts in the interior, and is affected by the chaotic phase only at
the boundaries.}
\label{fig:xmastree}
\end{figure}

In view of this discussion the space-time diagram can be separated in
two distinct phases: the ordered spatial period 4 and the chaotic
phase. The former can only be destroyed at the boundary or by spatial
defects (domains of period 5 or 6).

We verify this observation by two simulations. First,
the entire chain is initially set into period-4 state with the
period formed by two neighboring sites with $x_n(0)=x^{(1)}$, $n=1,2$ 
and the next two sites with $x_n(0)=x^{(2)}$, $n=3,4$.
The values $x^{(1)}$ and $x^{(2)}$ are taken to be equal to those
corresponding to the two peaks in $\rho(x)$ in
Fig.~\ref{fig:pert_distr} (f) for $r=3.83$ and $N=1000$.
Simulations starting from this initial conditions remain frozen in
this spatial period-4 and temporal period-2 state
indefinitely. Although metastable, this state requires the presence of
defects to be destroyed. 

This can be checked in a second test. We first iterate a 1000-site
lattice at $r=3.83$, for $10^6$ time-steps in order to eliminate any
transient effects. At that point, at once, half of the chain sites are
set into the same spatial period-4 state as above. As shown in
Fig.~\ref{fig:xmastree}, this phase then disappears gradually, over
about 30,000 time steps, invaded from the edges by the chaotic phase.

\section{Discussion}
\label{sec:discussion}

The previous sections have shown that the dynamics of the network can
be understood in terms of a competition between two dynamical regimes:
a stable period-two orbit and a fully chaotic state.
This can be deduced from Figs.~\ref{fig:bdonepconst} and
\ref{fig:pert_distr}: when the neighboring sites are in opposite
bands, their contribution shifts the bifurcation diagram into a
period-two regime. A spatial period 4 is therefore the basic stable
motif with this set of parameters. 

As mentioned above, we cannot reconstruct a stretched exponential
dynamics simply using the natural invariant density inserted into 
Eq.~(\ref{eq:pert_map}). 
We have found that the only way to obtain a stretched exponential
trap-time distribution in this situation is to impose a stretched
exponential trap-time distribution on $a(t)$ for the values of $a$
within the peak region in $\rho(a)$ (see Fig.~\ref{fig:pert_distr}).
More interestingly, connecting a single site, with a unidirectional
coupling, to two sites selected at random on a lattice, is sufficient
to induce a stretched exponential trap distribution on this single
site, albeit with a larger $\beta$ than that for the elements which
belong to the lattice.

Thus once an external perturbation, following a stretched exponential
dynamics, is imposed on a chaotic element, the latter will immediately
adopt a similar dynamics.
However, stretched exponential distributions cannot be observed in a
self-organized process without spatial organization.
For example, connecting the nodes of a balanced binary tree 
unidirectionally towards its root results in the densities $\rho(x)$ 
and $\rho(a)$ similar to those depicted in Fig.~\ref{fig:pert_distr}, 
(e) and (f) already for two levels in the tree. 
Nevertheless, the trap-time distribution measured at the root node 
is a pure exponential.
This suggests that the stretched exponential distributions requires
some spatial organization allowing time-limited spatial period-4
structures to occur and to be stabilized.

The nature of this spatial organization is somewhat
paradoxical. Correlations are very short range and the width of these
domain is only weakly varying with the stretching
parameter. While traps at $r=3.83$ are up to 250 times longer than
those at $r=3.8888$, their width is only 4 to 5 times larger, as
displayed in
Fig.~\ref{fig:bin_map}.
This decoupling between spatial and temporal
correlation is reminiscent of dynamical heterogeneities observed in in
glass-forming systems
\cite{Garrahan_2002,VollmayrLee_2002,Dzugutov_2001}.
Associating a ``temperature'' with the control parameter $r$, we see
that the spatial size of the traps increases with decreasing $r$
without diverging.
In the same way, the size of dynamical heterogeneities increases and
the stretching exponent $\beta$ decreases under cooling in
glass-forming systems.

As we showed also, once in a perfect spatial period-4 regime, the system
will never become chaotic. The finite life-time arises from defects in
the period-4 phase or from the chaotic boundaries. 
This behavior is similar to what is seen in the spin-system models
with effective constraints on the dynamics \cite{Garrahan_2002} and
the state-space partition model \cite{Nadler_96}.
Interestingly, both models show a stretched exponential decay of
certain statistical quantities.
The distinctive feature of the present model, Sect.~\ref{sec:model} is
that the effective dynamical constraints arise in the course of the
evolution described by the dynamical equation~(\ref{cml}), unlike the
above cited models where the fixed constrains are imposed on the
neighboring sites and the dynamics is due to a Monte Carlo procedure.
%
%We plan to provide a detailed probability theoretical analysis of the
%dynamics of the present model in a future study.

\section{Conclusion}
\label{sec:conclusion}

The coupled array of chaotic oscillators presented here has a number
of properties that make it an important model.
As was shown above and in Ref.~\onlinecite{Hunt_2002}, it is a
faithful representation of the dynamics of an experimental set-up of
coupled diode-resonators in chaotic regime characterized by a
stretched exponential distribution of trap time.
Moreover, the low cost associated with solving numerically the model
of Eq.~(\ref{cml}) allows us to study the system on time scale
unreachable with atomic models; it takes less than a day, on a fast
processor, to iterate a lattice of 1000 sites over $10^{9}$
steps. With traps extending to more than 30,000 times the basic time
step, long simulations are absolutely necessary to establish the
nature of the dynamics in these systems.

Results of these simulations demonstrate that the stretched
exponential distributions arise from the competition between a chaotic
and a period-two regimes. 
The stretched-exponential requires some spatial organization to appear
but does no imply diverging length scale as the traps become longer
and longer: the period-4 structure is sufficient to stabilize a site
onto a periodic orbit and the space of the spatial correlation is only
weakly related to the length of the traps appearing in these systems.

There are strong similarities between this system and the
configurational glasses and we suggest that the understanding gained
here could be extended directly to these important materials. We are
currently pursuing this avenue of research.

\section*{Acknowledgments}

We thank C.~Godr{\`e}che and I.A.~Campbell for valuable comments. This work in
supported in part by the Office of Naval Research (ERH),  and by the Natural
Sciences and Engineering Council of Canada and the NATEQ fund of Qu\'ebec (NM
and SIS). Most of the calculations were one on the computers of the R\'eseau
qu\'eb\'ecois de calcul de haute performance (RQCHP). NM is a Cottrell Scholar
of the Research Corporation.

%\bibliographystyle{apsrev}
%\bibliographystyle{plain}
%\bibliography{archive}

\end{document}